\documentclass[10pt]{article}
\usepackage{multicol}
\usepackage{graphicx}
\usepackage{amsmath}
\usepackage[a4paper]{geometry}
\usepackage{hyperref}

\begin{document}
\begin{center}
{\Large \bf Initial, effective, and kinetic freeze-out
temperatures from transverse momentum spectra in high energy
proton(deuteron)-nucleus and nucleus-nucleus collisions}

\vskip.75cm

Muhammad Waqas, Fu-Hu Liu{\footnote{E-mail: fuhuliu@163.com;
fuhuliu@sxu.edu.cn}}

{\small\it Institute of Theoretical Physics and State Key
Laboratory of Quantum Optics and Quantum Optics Devices,\\ Shanxi
University, Taiyuan, Shanxi 030006, China}
\end{center}

\vskip.5cm

{\bf Abstract:} The transverse momentum spectra of charged
particles produced in proton(deuteron)-nucleus and nucleus-nucleus
collisions at high energies are analyzed by the Hagedorn thermal
model and the standard distribution in terms of multi-component.
The experimental data measured in central and peripheral gold-gold
(Au-Au) and deuteron-gold ($d$-Au) collisions by the PHENIX
Collaboration at the Relativistic Heavy Ion Collider (RHIC), as
well as in central and peripheral lead-lead (Pb-Pb) and
proton-lead ($p$-Pb) collisions by the ALICE Collaboration at the
Large Hadron Collider (LHC) are fitted by the two models. The
initial, effective, and kinetic freeze-out temperatures are then
extracted from the fitting to the transverse momentum spectra. It
is shown that the initial temperature is larger than the effective
temperature, and the effective temperature is larger than the
kinetic freeze-out temperature. The three types of temperatures in
central collisions are comparable with those in peripheral
collisions, and those at the LHC are comparable with those at the
RHIC.
\\

{\bf Keywords:} Transverse momentum spectra, initial temperature,
kinetic freeze-out temperature, nucleus-nucleus collisions
\\

{\bf PACS:} 14.40.-n, 14.20.-c, 24.10.Pa

\vskip1.0cm

\begin{multicols}{2}

{\section{Introduction}}

The concept of temperature used in thermodynamics and statistical
mechanics~\cite{1} is also used in subatomic physics due to
similar thermal and statistical property. At least the initial,
chemical freeze-out, thermal or kinetic freeze-out, and effective
temperatures are used in high energy collisions. The initial
temperature describes the excitation degree of interacting system
at the initial stage of collisions. The chemical freeze-out
temperature describes the excitation degree of interacting system
at the stage of chemical freeze-out, where the ratios of different
types of particles are no longer changed. The kinetic freeze-out
temperature describes the excitation degree of interacting system
at the stage of kinetic freeze-out, where the transverse momentum
spectra are no longer changed. The effective temperature describes
the sum of the excitation degree of interacting system and the
effect of transverse flow at the stage of kinetic freeze-out,
where the transverse flow resulted from the impact and squeeze
reflects the hydrodynamic expansion of interacting systems.

The initial temperature has less studies in the community due to
undefined method, though it should be based on the particle
spectra. The chemical freeze-out temperature can be obtained from
the particle ratios~\cite{2,3,4,5}. The kinetic freeze-out
temperature can be obtained from the transverse momentum spectra,
in the case of getting rid of the transverse flow effect and
leaving only the contribution of random thermal
motion~\cite{6,7,8,9,10}. If the temperature extracted from the
transverse momentum spectra also contains the contribution of
transverse flow, this type of temperature is not the kinetic
freeze-out temperature, but the effective temperature called by
us. The initial, chemical freeze-out, and kinetic freeze-out
temperatures are ``real" temperatures. The effective temperature
is not a ``real" temperature.

Generally, the initial stage of collisions happens earlier than
the stages of chemical and kinetic freeze-outs. The initial
temperature should be the largest among the three ``real"
temperatures. The initial stage of collisions is also possible to
be simultaneous with the stage of chemical freeze-out. This also
results possibly in the initial temperature to be equal to the
chemical freeze-out temperature. Similarly, the stage of chemical
freeze-out is simultaneous with or earlier than that of kinetic
freeze-out. This results in the chemical freeze-out temperature to
be equal to or larger than the kinetic freeze-out temperature.
Because of two contributions, the effective temperature is often
larger than the kinetic freeze-out temperature. In most cases, one
has the initial temperature, chemical freeze-out temperature,
effective temperature, and kinetic freeze-out temperature to
reduce in turn.

Due to the definiteness of particle ratios at given condition such
as at mid-rapidity in central collisions at given collision
energy, the chemical freeze-out temperature can be measured
correspondingly by the thermal and statistical
model~\cite{2,3,4,5}. Comparatively, the initial, effective, and
kinetic freeze-out temperatures are model dependent due to the
fact that different functions or distributions are used to fit the
transverse momentum spectra~\cite{6,7,8,9,10}. At the same time,
the limiting value of chemical freeze-out temperature at very high
energy is knowable, and the limiting values of initial, effective,
and kinetic freeze-out temperatures if available at high energy
are difficult to predict due to their erratic trends.

It is unquestioned that the mentioned temperatures are related to
study the phase diagram in the temperature related spaces, such as
the temperatures as functions of transverse flow velocity, baryon
chemical potential, collision centrality, system size, particle
rapidity, and collision energy. To study systematically these
temperature related functions are a huge project, in particular
for the dependences of temperatures on collision energy. As a
beginning of the project, the dependences of temperatures on
collision centrality can be studied. Naturally, the studies of
dependences of temperatures on these quantities are useful for us
to understand the mechanisms of multi-particle production and
system evolution. Although the initial (kinetic freeze-out)
temperature is model dependent in most cases, we hope to use a
model independent method to obtain it. The effective temperature
is definitely model dependent, we shall not pay much attention to
it.

It should be noted that there are different methods~\cite{30} to
extract the kinetic freeze-out temperature from the transverse
momentum spectra in high energy collisions~\cite{10}. These
methods include, but are not limited to, the blast-wave model with
the Boltzmann statistics~\cite{6,7,8}, the blast-wave model with
the Tsallis statistics ~\cite{9,30a,30b}, an alternative
method~\cite{7,30c,30d,30e,30f} which describes the transverse
momentum spectra by using a Boltzmann distribution~\cite{17} and
the intercept in the effective temperature versus particle mass is
regarded as the kinetic freeze-out temperature, and the same
alternative method but using a Tsallis distribution~\cite{17,18}.
Our recent work~\cite{30} shows that the four methods are
harmonious in the dependences of kinetic freeze-out temperature on
collision centrality and energy, though the absolute values are
different in some cases.

In this paper, the transverse momentum spectra of positively and
negatively charged pions ($\pi^+$ and $\pi^-$), positively and
negatively charged kaons ($K^+$ and $K^-$), and protons ($p$) and
antiprotons ($\bar p$) produced in central and peripheral
gold-gold (Au-Au) and deuteron-gold ($d$-Au) collisions at the
Relativistic Heavy Ion Collider (RHIC)~\cite{11,12}, as well as in
central and peripheral lead-lead (Pb-Pb) and proton-lead ($p$-Pb)
collisions at the Large Hadron Collider (LHC)~\cite{13,14,15} are
fitted by the Hagedorn thermal model~\cite{16} and Boltzmann
distribution~\cite{17} in terms of multi-component which is
harmonious with a multisource thermal model used in our previous
work~\cite{10}. The related temperatures which include the
initial, effective, and kinetic freeze-out temperatures are then
extracted from the fittings. In particular, the alternative method
is used to extract the kinetic freeze-out temperature.

The remainder of this paper is structured as follows. The
formalism and method are shortly described in Section 2. Results
and discussion are given in Section 3. In Section 4, we summarize
our main observations and conclusions.
\\

{\section{Formalism and method}}

There are two main processes of multi-particle productions in high
energy collisions, namely the soft excitation and hard scattering
processes. For the two main processes, one can use different
formalisms to describe the transverse momentum spectra of charged
particles. Generally, in the descriptions, the soft excitation
process has many choices of formalisms, and the hard scattering
process has very limited choices of formalisms.

For the soft excitation process, the choices include, but are not
limited to, the Hagedorn thermal model (the statistical-bootstrap
model)~\cite{16}, the (multi-)standard distribution~\cite{17}, the
Tsallis and related distributions with various
formalisms~\cite{18}, the blast-wave model with Boltzmann
statistics~\cite{6,7,8}, the blast-wave model with Tsallis
statistics~\cite{9}, etc. To focus our attention on the effective
temperature and then the initial and kinetic freeze-out
temperatures according to the transverse momentum spectra, we use
only the Hagedorn thermal model~\cite{16} and the standard
distribution~\cite{17} in terms of multi-component as examples in
the present work. The chemical freeze-out temperature needs
systematically the particle ratios, which is beyond the focus of
our attention and is not discussed in the present work.

According to ref.~\cite{16}, the Hagedorn thermal model results in
the transverse momentum ($p_T$) distribution, in terms of the
probability density function, at the mid-rapidity (the rapidity of
the center-of-mass of interacting system in the center-of-mass
system) to be
\begin{align}
f_1(p_T)=&\frac{1}{N}\frac{dN}{dp_T}=C_1p_T\sqrt{p_T^2+m_0^2} \nonumber\\
&\times\sum_{n=1}^{\infty}\left(-S\right)^{n+1}
K_1\left(n\frac{\sqrt{p_T^2+m_0^2}}{T_1} \right),
\end{align}
where $N$, $C_1$, $m_0$, $n$, $K_1$, and $T_1$ denote the particle
number, normalization constant, particle rest mass, item number in
the summation, modified Bessel function of the second kind, and
effective temperature, respectively, and $S=+1$ and $-1$ for
fermions and bosons respectively. Generally, to obtain suitable
$p_T$ distributions, $n$ is taken to be 1--10 for pions, 1--5 for
kaons, and 1 for nucleons and hyperons~\cite{16}. In Eq. (1), the
chemical potential is not considered due to very small value and
effect at RHIC and LHC energies.

For particles distributed in a wide rapidity ($y$) range from the
minimum rapidity $y_{\min}$ to the maximum rapidity $y_{\max}$,
and the mid-rapidity is in $[y_{\min}, y_{\max}]$, we have the
$p_T$ distribution to be
\begin{align}
f_1(p_T)=& C_1p_T\sqrt{p_T^2+m_0^2} \int_{y_{\min}}^{y_{\max}} \cosh y \nonumber\\
&\times\sum_{n=1}^{\infty}\left(-S\right)^{n+1}
K_1\left(n\frac{\sqrt{p_T^2+m_0^2}\cosh y}{T_1}\right)dy,
\end{align}
where the normalization constant $C_1$ in Eq. (2) may be different
from that in Eq. (1). Comparing with Eq. (1) which is the
particular case of Eq. (2) with $y=0$, Eq. (2) is more suitable to
the particles distributed in a wide rapidity range. In some cases,
the mid-rapidity is not in $[y_{\min}, y_{\max}]$. Then, we may
transform the rapidity range by adding or subtracting a rapidity
shift to cover the mid-rapidity, so that we can exclude the
contribution of directional movement of the emission source.

In some cases, the Hagedorn thermal model is not enough to
describe the spectra in low $p_T$ region contributed by the soft
process. A two-, three-, or even multi-component Hagedorn thermal
model is needed. We have the multi-component model to be
\begin{align}
f_1(p_T)=& \sum_{i=1}^{l}
k_{1i} C_{1i} p_T\sqrt{p_T^2+m_0^2} \nonumber\\
&\times\sum_{n=1}^{\infty}\left(-S\right)^{n+1}
K_1\left(n\frac{\sqrt{p_T^2+m_0^2}}{T_{1i}} \right),
\end{align}
or
\begin{align}
f_1(p_T)=& \sum_{i=1}^{l} k_{1i} C_{1i} p_T \sqrt{p_T^2+m_0^2}
\int_{y_{\min}}^{y_{\max}} \cosh y \nonumber\\
&\times\sum_{n=1}^{\infty}\left(-S\right)^{n+1}
K_1\left(n\frac{\sqrt{p_T^2+m_0^2}\cosh y}{T_{1i}} \right)dy,
\end{align}
where $l$ denotes the number of components, and $k_{1i}$,
$C_{1i}$, and $T_{1i}$ denote the contribution fraction,
normalization constant, and effective temperature corresponding to
the $i$-th component respectively. In particular, the
normalization results in $\sum_{i=1}^l k_{1i}=1$, and the
effective temperature averaged by weighting different components
is $T_1=\sum_{i=1}^l k_{1i}T_{1i}$. Generally, $l\leq3$.

The standard distribution is a joint name of the Boltzmann,
Fermi-Dirac, and Bose-Einstein distributions which base on the
Boltzmann, Fermi-Dirac, and Bose-Einstein statistics,
respectively, which correspond to $S=0$, $+1$, and $-1$,
respectively~\cite{17}. At the mid-rapidity or in the rapidity
range $[y_{\min},y_{\max}]$ which covers the mid-rapidity, the
standard distribution is
\begin{align}
f_2(p_T)=& C_2 p_T \sqrt{p_T^2+m_0^2} \nonumber\\
&\times \left[ \exp\left( \frac{\sqrt{p_T^2+m_0^2}}{T_2} \right)+S
\right]^{-1}
\end{align}
or
\begin{align}
f_2(p_T)=& C_2 p_T \sqrt{p_T^2+m_0^2} \int_{y_{\min}}^{y_{\max}} \cosh y \nonumber\\
&\times \left[ \exp\left( \frac{\sqrt{p_T^2+m_0^2}\cosh y}{T_2}
\right)+S \right]^{-1}dy,
\end{align}
where $C_2$ and $T_2$ are the normalization constant and effective
temperature respectively. If the rapidity range
$[y_{\min},y_{\max}]$ does not cover the mid-rapidity, we need to
shift it to cover the mid-rapidity.

In some cases, the standard distribution is not enough to describe
the spectra in low $p_T$ region contributed by the soft process. A
two-, three-, or even multi-component standard distribution is
needed. We have the multi-component standard distribution to be
\begin{align}
f_2(p_T)=&\sum_{i=1}^{l} k_{2i} C_{2i} p_T \sqrt{p_T^2+m_0^2} \nonumber\\
&\times \left[ \exp\left( \frac{\sqrt{p_T^2+m_0^2}}{T_{2i}}
\right)+S \right]^{-1},
\end{align}
or
\begin{align}
f_2(p_T)=&\sum_{i=1}^{l} k_{2i} C_{2i} p_T \sqrt{p_T^2+m_0^2}
\int_{y_{\min}}^{y_{\max}} \cosh y \nonumber\\
&\times \left[ \exp\left( \frac{\sqrt{p_T^2+m_0^2}\cosh y}{T_{2i}}
\right)+S \right]^{-1}dy,
\end{align}
where $k_{2i}$, $C_{2i}$, and $T_{2i}$ denote the contribution
fraction, normalization constant, and effective temperature
corresponding to the $i$-th component respectively. In particular,
the normalization results in $\sum_{i=1}^l k_{2i}=1$, and the
effective temperature averaged by weighting different components
is $T_2=\sum_{i=1}^l k_{2i}T_{2i}$.

It should be noted that the rapidity $y$ used in Eqs. (4) and (8)
are for particles, but not for fireballs. If we study rapidity
spectra, we should use these fireballs at different rapidities
$y_x$. In the case of studying $p_T$ spectra, $y_x$ should be
directly shifted to 0 so that the kinetic energy of directional
movement of fireballs can be removed from temperature which is
contributed by thermal motion, but not directional movement, of
particles.

For a not too wide $p_T$ spectrum, the above equations such as
Eqs. (3) or (4) and (7) or (8) can be used to describe the $p_T$
spectrum and to extract the effective temperature. For a wide
$p_T$ spectrum, we have to consider the contribution of hard
scattering process. In some cases, although the $p_T$ spectrum is
wide, the contribution of hard scattering process in high-$p_T$
regions is negligible. We can only concern the contribution of
soft excitation process in low-$p_T$ region. If the spectrum of
high-$p_T$ region is non-negligible, we can use an inverse
power-law, i.e. the Hagedorn function~\cite{16}
\begin{align}
f_H(p_T) =Ap_T \bigg( 1+\frac{p_T}{ p_0} \bigg)^{-n}
\end{align}
to describe the contribution of hard scattering process, where
$p_0$ and $n$ are free parameters, and $A$ is the normalization
constant related to the free parameters. The inverse power-law is
obtained from the quantum chromodynamics (QCD)
calculus~\cite{19,20,21} and has three
revisions~\cite{22,23,24,25,26,27,28} which will not be discussed
further in the present work.

In the case of considering both the contributions of soft
excitation and hard scattering processes, the experimental $p_T$
spectrum distributed in a wide range can be described by a
superposition, i.e.
\begin{align}
f_0(p_T)=k_0f_S(p_T)+(1-k_0)f_H(p_T),
\end{align}
where $k_0$ denotes the contribution fraction of the soft
excitation process and $f_S(p_T)$ denotes one of Eqs. (3) or (4)
and (7) or (8). According to Hagedorn's model~\cite{16}, we may
also use the usual step function
\begin{align}
f_0(p_T)=A_1 \theta(p_1-p_T) f_S(p_T) + A_2
\theta(p_T-p_1)f_H(p_T),
\end{align}
to superpose the two functions, where $A_1$ and $A_2$ are
constants which result in the two components to be equal to each
other at $p_T=p_1 \approx 2\sim3$ GeV/$c$.

Eqs. (10) and (11) are two different superpositions. In Eq. (10),
the soft component contributes from 0 up to $2\sim3$ GeV/$c$ or a
little more, and the hard component contributes in the whole $p_T$
range. The main contributor in the low-$p_T$ range is the soft
component and the only contributor in the high-$p_T$ range is the
hard component. In Eq. (11), the soft component contributes from 0
up to $p_1$, and the hard component contributes from $p_1$ up to
the maximum. There is no mixed range for the two components,
though the curve is possibly not smooth at their boundary $p_1$.
We shall use only the first component in Eq. (11) due to not too
wide $p_T$ range studied in the present work. In particular,
$f_S(p_T)$ is exactly Eqs. (4) and (8) respectively, in which
$l=1$, 2, or 3 for different sets of data.
\\

{\section{Results and discussion}}

Figure 1 presents the transverse momentum spectra, $(1/2\pi p_T)
d^2N/dp_Tdy$, of (a)-(c) $\pi^+$, $K^+$, and $p$, as well as
(b)-(d) $\pi^-$, $K^-$, and $\bar p$ produced in (a)-(b) 0--5\%
and (c)-(d) 60--92\% Au-Au collisions at center-of-mass energy per
nucleon pair $\sqrt{s_{NN}}=200$ GeV. The squares, circles, and
triangles represent respectively the experimental data of $\pi^+$
($\pi^-$), $K^+$ ($K^-$), and $p$ ($\bar p$) measured by the
PHENIX Collaboration in the pseudorapidity range
$|\eta|<0.35$~\cite{11}. The solid and dotted curves are our
results fitted by the first component in Eq. (11) through Eqs. (4)
and (8) respectively, where we have coded the equation ourselves
by Matlab which has intrinsic functions to perform calculations
for Bessel and modified Bessel functions. The values of related
parameters are listed in Tables 1 and 2 with the value of $\chi^2$
and the number of degree of freedom (dof) in terms of
$\chi^2$/dof. The ratios of data/fit corresponding to panels
(a)--(d) are presented by panels (a$^*$)--(d$^*$) respectively,
where the closed squares, circles, and triangles are the results
for $\pi^+$ ($\pi^-$), $K^+$ ($K^-$), and $p$ ($\bar p$) due to
Eqs. (4), and the open squares, circles, and triangles are the
results for $\pi^+$ ($\pi^-$), $K^+$ ($K^-$), and $p$ ($\bar p$)
due to Eqs. (8). One can see that the two models fit well the
trends of the experimental data in low-$p_T$ region in Au-Au
collisions at the top RHIC energy.

Figure 2 is the same as Fig. 1, but it shows the the spectra in
(a)-(b) 0--20\% and (c)-(d) 60--88\% $d$-Au collisions at
$\sqrt{s_{NN}}=200$ GeV. The symbols represent the experimental
data measured by the PHENIX Collaboration in
$|\eta|<0.35$~\cite{12}. The curves are our fitted results and the
related parameters are listed in Tables 1 and 2 with $\chi^2$/dof.
One can see that the two models fit well the trends of the
experimental data in low-$p_T$ region in $d$-Au collisions at the
top RHIC energy.

\begin{figure*}[htbp]
\begin{center}
\includegraphics[width=13.0cm]{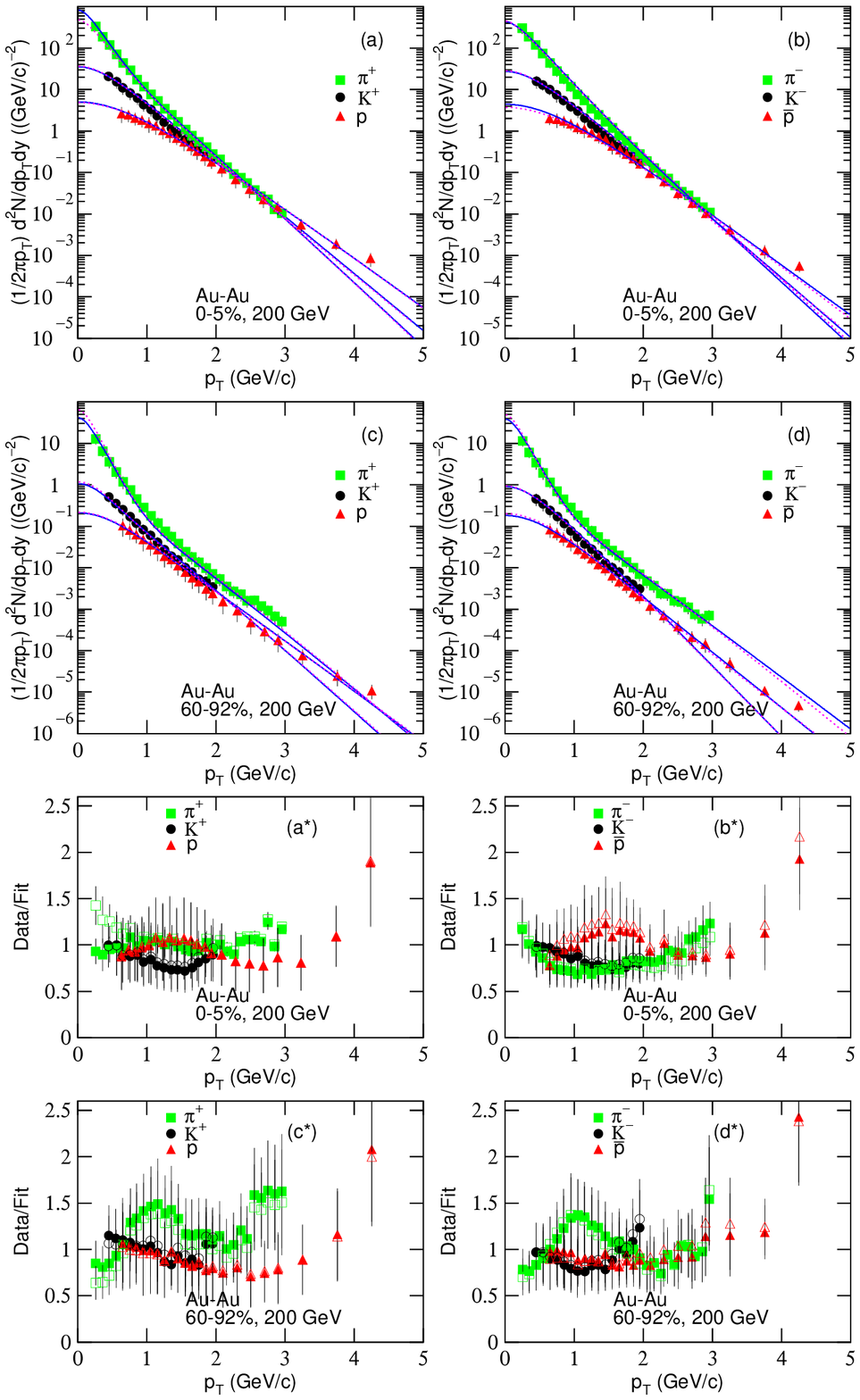}
\end{center}
{\small Fig. 1. Transverse momentum spectra of (a)-(c) $\pi^+$,
$K^+$, and $p$, as well as (b)-(d) $\pi^-$, $K^-$, and $\bar p$
produced in (a)-(b) 0--5\% and (c)-(d) 60--92\% Au-Au collisions
at $\sqrt{s_{NN}}=200$ GeV. The symbols represent the experimental
data measured by the PHENIX Collaboration in
$|\eta|<0.35$~\cite{11}. The solid and dotted curves are our
results fitted by Eqs. (4) and (8) respectively. The ratios of
data/fit corresponding to panels (a)--(d) are presented by panels
(a$^*$)--(d$^*$) respectively, where the closed and open symbols
are the results due to Eqs. (4) and (8) respectively.}
\end{figure*}

\begin{table*}
{\small Table 1. Values of free parameters ($T_{11}$, $T_{12}$,
$T_{13}$ if available, $k_{11}$, and $k_{12}$ if available),
normalization constant ($N_0$), and $\chi^2$/dof corresponding to
the solid curves in Figs. 1--4.} \vspace{-.30cm} {\tiny
\begin{center}
\begin{tabular}{cccccccccccc}\\ \hline\hline
Figure & Centrality & Particle & $T_{11}$ (GeV) & $T_{12}$ (GeV) &
$T_{13}$ (GeV) &
$k_{11}$ & $k_{12} $& $N_0$ & $\chi^2$/dof\\
\hline
Fig. 1   & 0--5\%   & $\pi^+$  & $0.267\pm0.009$ & $0.544\pm0.006$ & $-$             & $0.70\pm0.05$ & $-$           & $50.95\pm4.00$  & $315/25$\\
Au-Au    &          & $K^+$    & $0.460\pm0.006$ & $0.605\pm0.007$ & $0.614\pm0.005$ & $0.46\pm0.08$ & $0.20\pm0.04$ & $8.70\pm0.70$   & $9/13$\\
200 GeV  &          & $p$      & $0.685\pm0.005$ & $0.686\pm0.005$ & $-$             & $0.55\pm0.02$ & $-$           & $2.51\pm0.20$   & $5/19$\\
         &          & $\pi^-$  & $0.378\pm0.006$ & $0.551\pm0.008$ & $-$             & $0.70\pm0.05$ & $-$           & $44.95\pm5.00$  & $39/25$\\
         &          & $K^-$    & $0.437\pm0.008$ & $0.584\pm0.008$ & $-$             & $0.16\pm0.06$ & $-$           & $7.20\pm0.60$   & $10/13$\\
         &          & $\bar p$ & $0.647\pm0.005$ & $0.773\pm0.006$ & $-$             & $0.92\pm0.16$ & $-$           & $2.12\pm0.14$   & $7/19$\\
\cline{2-10}
         & 60--92\% & $\pi^+$  & $0.238\pm0.008$ & $0.636\pm0.006$ & $0.715\pm0.007$ & $0.90\pm0.05$ & $0.11\pm0.01$ & $1.83\pm0.18$   & $17/25$\\
         &          & $K^+$    & $0.260\pm0.006$ & $0.560\pm0.008$ & $-$             & $0.35\pm0.09$ & $-$           & $0.17\pm0.02$   & $2/13$\\
         &          & $p$      & $0.395\pm0.005$ & $0.666\pm0.006$ & $-$             & $0.48\pm0.04$ & $-$           & $0.070\pm0.005$ & $6/19$\\
         &          & $\pi^-$  & $0.235\pm0.005$ & $0.323\pm0.008$ & $0.669\pm0.007$ & $0.80\pm0.07$ & $0.10\pm0.01$ & $1.74\pm0.16$   & $11/25$\\
         &          & $K^-$    & $0.375\pm0.008$ & $0.490\pm0.005$ & $ -$            & $0.24\pm0.08$ & $-$           & $0.18\pm0.02$   & $5/13$\\
         &          & $\bar p$ & $0.418\pm0.014$ & $0.624\pm0.005$ & $-$             & $0.58\pm0.04$ & $-$           & $0.060\pm0.010$ & $6/19$\\
\hline
Fig. 2   & 0--20\%  & $\pi^+$  & $0.316\pm0.007$ & $0.515\pm0.005$ & $0.978\pm0.008$ & $0.60\pm0.05$ & $0.36\pm0.04$ & $0.99\pm0.05$   & $6/21$\\
$d$-Au   &          & $K^+$    & $0.348\pm0.008$ & $0.478\pm0.008$ & $0.864\pm0.005$ & $0.42\pm0.03$ & $0.29\pm0.03$ & $0.13\pm0.01$   & $1/18$\\
200 GeV  &          & $p$      & $0.625\pm0.005$ & $0.752\pm0.004$ & $-$             & $0.56\pm0.08$ & $-$           & $0.055\pm0.004$ & $9/21$\\
         &          & $\pi^-$  & $0.290\pm0.008$ & $0.506\pm0.006$ & $1.100\pm0.002$ & $0.51\pm0.03$ & $0.46\pm0.04$ & $0.99\pm0.06$   & $5/21$\\
         &          & $K^-$    & $0.378\pm0.007$ & $0.393\pm0.012$ & $0.852\pm0.008$ & $0.40\pm0.03$ & $0.30\pm0.03$ & $0.13\pm0.01$   & $2/18$\\
         &          &$\bar p$  & $0.490\pm0.008$ & $0.612\pm0.007$ & $0.712\pm0.008$ & $0.20\pm0.06$ & $0.45\pm0.07$ & $0.050\pm0.004$ & $15/21$\\
\cline{2-10}
         & 60--88\% & $\pi^+$  & $0.296\pm0.008$ & $0.666\pm0.003$ & $0.670\pm0.005$ & $0.74\pm0.04$ & $0.74\pm0.10$ & $0.21\pm0.02$   & $23/21$\\
         &          & $K^+$    & $0.307\pm0.008$ & $0.324\pm0.014$ & $0.778\pm0.006$ & $0.42\pm0.02$ & $0.30\pm0.03$ & $0.036\pm0.003$ & $2/18$\\
         &          & $p$      & $0.410\pm0.009$ & $0.788\pm0.006$ & $-$             & $0.75\pm0.02$ & $-$           & $0.015\pm0.001$ & $12/21$\\
         &          & $\pi^-$  & $0.258\pm0.008$ & $0.650\pm0.007$ & $0.765\pm0.004$ & $0.70\pm0.01$ & $0.30\pm0.01$ & $0.21\pm0.02$   & $21/21$\\
         &          & $K^-$    & $0.318\pm0.010$ & $0.450\pm0.011$ & $0.885\pm0.005$ & $0.38\pm0.03$ & $0.32\pm0.02$ & $0.035\pm0.002$ & $4/18$\\
         &          & $\bar p$ & $0.160\pm0.010$ & $0.280\pm0.010$ & $0.600\pm0.005$ & $0.08\pm0.01$ & $0.05\pm0.01$ & $0.011\pm0.001$ & $19/21$\\
\hline
Fig. 3   & 0--5\%   & $\pi^++\pi^-$ & $0.236\pm0.012$ & $0.597\pm0.005$ & $-$             & $0.41\pm0.04$ & $-$           & $240.00\pm15.00$& $28/38$\\
Pb-Pb    &          & $K^++K^-$     & $0.674\pm0.008$ & $0.684\pm0.009$ & $-$             & $0.30\pm0.05$ & $-$           & $36.00\pm1.70$  & $30/33$\\
2.76 TeV &          & $p+\bar p$    & $0.970\pm0.009$ & $1.050\pm0.001$ & $1.150\pm0.001$ & $0.65\pm0.05$ & $0.84\pm0.06$ & $9.70\pm0.33$   & $9/34$\\
\cline{2-10}
         & 70--80\% & $\pi^++\pi^-$ & $0.272\pm0.013$ & $0.683\pm0.006$ & $-$             & $0.74\pm0.02$ & $-$           & $5.60\pm0.29$   & $458/38$\\
         &          & $K^++K^-$     & $0.183\pm0.013$ & $0.620\pm0.005$ & $-$             & $0.16\pm0.04$ & $-$           & $0.81\pm0.04$   & $14/33$\\
         &          & $p+\bar p$    & $0.393\pm0.016$ & $0.401\pm0.012$ & $0.730\pm0.006$ & $0.01\pm0.01$ & $0.02\pm0.01$ & $0.30\pm0.01$   & $9/34$\\
\hline
Fig. 4   & 0--5\%   & $\pi^++\pi^-$ & $0.215\pm0.014$ & $0.600\pm0.003$ & $0.729\pm0.007$ & $0.54\pm0.04$ & $0.09\pm0.03$ & $6.73\pm0.33$   & $20/38$\\
$p$-Pb   &          & $K^++K^-$     & $0.260\pm0.014$ & $0.294\pm0.014$ & $0.796\pm0.006$ & $0.10\pm0.02$ & $0.12\pm0.04$ & $0.95\pm0.04$   & $5/28$\\
5.02 TeV &          & $p+\bar p$    & $0.878\pm0.010$ & $0.930\pm0.009$ & $-$             & $0.56\pm0.16$ & $-$           & $0.37\pm0.02$   & $19/36$\\
\cline{2-10}
         & 60--80\% & $\pi^++\pi^-$ & $0.203\pm0.013$ & $0.365\pm0.009$ & $0.750\pm0.006$ & $0.49\pm0.02$ & $0.29\pm0.02$ & $1.65\pm0.10$   & $41/38$\\
         &          & $K^++K^-$     & $0.219\pm0.008$ & $0.327\pm0.009$ & $0.693\pm0.005$ & $0.21\pm0.04$ & $0.13\pm0.01$ & $0.19\pm0.01$   & $14/28$\\
         &          & $p+\bar p$    & $0.434\pm0.016$ & $0.860\pm0.011$ & $0.971\pm0.007$ & $0.54\pm0.03$ & $0.15\pm0.05$ & $0.19\pm0.01$   & $17/36$\\
\hline\hline
\end{tabular}
\end{center}}
\end{table*}

\begin{table*}
{\small Table 2. Values of free parameters ($T_{21}$, $T_{22}$,
$T_{23}$ if available, $k_{21}$, and $k_{22}$ if available),
normalization constant ($N_0$), and $\chi^2$/dof corresponding to
the dashed curves in Figs. 1--4.} \vspace{-.30cm} {\tiny
\begin{center}
\begin{tabular}{ccccccccccc}\\ \hline\hline
Figure & Centrality & Particle & $T_{21}$ (GeV) & $T_{22}$ (GeV) &
$T_{23}$ (GeV) &
$k_{21}$ & $k_{22} $& $N_0$ & $\chi^2$/dof\\
\hline
Fig. 1  & 0--5\%   & $\pi^+$  & $0.148\pm0.008$ & $0.261\pm0.004$ & $-$             & $0.08\pm0.01$ & $-$           & $50.95\pm5.00$  & $404/25$\\
Au-Au   &          & $K^+$    & $0.225\pm0.007$ & $0.299\pm0.005$ & $-$             & $0.13\pm0.06$ & $-$           & $8.73\pm0.60$   & $5/13$\\
200 GeV &          & $p$      & $0.281\pm0.007$ & $0.332\pm0.003$ & $-$             & $0.02\pm0.05$ & $-$           & $2.52\pm0.15$   & $4/19$\\
        &          & $\pi^-$  & $0.179\pm0.008$ & $0.191\pm0.009$ & $0.266\pm0.003$ & $0.60\pm0.02$ & $0.15\pm0.01$ & $45.00\pm4.70$  & $44/25$\\
        &          & $K^-$    & $0.235\pm0.005$ & $0.297\pm0.002$ & $-$             & $0.80\pm0.01$ & $-$           & $7.20\pm0.53$   & $3/13$\\
        &          & $\bar p$ & $0.309\pm0.006$ & $0.318\pm0.007$ & $-$             & $0.45\pm0.06$ & $-$           & $2.12\pm0.17$   & $8/19$\\
\cline{2-10}
        & 60--92\% & $\pi^+$  & $0.102\pm0.008$ & $0.112\pm0.006$ & $0.292\pm0.006$ & $0.20\pm0.03$ & $0.39\pm0.03$ & $1.82\pm0.20$   & $15/25$\\
        &          & $K^+$    & $0.132\pm0.008$ & $0.272\pm0.006$ & $-$             & $0.35\pm0.04$ & $-$           & $0.18\pm0.02$   & $1/13$\\
        &          & $p$      & $0.190\pm0.008$ & $0.209\pm0.008$ & $0.325\pm0.005$ & $0.20\pm0.03$ & $0.21\pm0.03$ & $0.069\pm0.004$ & $5/19$\\
        &          & $\pi^-$  & $0.102\pm0.007$ & $0.122\pm0.008$ & $0.302\pm0.006$ & $0.16\pm0.03$ & $0.40\pm0.07$ & $1.74\pm0.16$   & $13/25$\\
        &          & $K^-$    & $0.188\pm0.008$ & $0.241\pm0.006$ & $-$             & $0.46\pm0.08$ & $-$           & $0.18\pm0.02$   & $2/13$\\
        &          & $\bar p$ & $0.204\pm0.007$ & $0.305\pm0.004$ & $0.373\pm0.007$ & $0.45\pm0.02$ & $0.49\pm0.02$ & $0.061\pm0.007$ & $7/19$\\
\hline
Fig. 2  & 0--20\%  & $\pi^+$  & $0.155\pm0.009$ & $0.247\pm0.005$ & $0.468\pm0.005$ & $0.58\pm0.04$ & $0.24\pm0.03$ & $0.97\pm0.06$   & $10/21$\\
$d$-Au  &          & $K^+$    & $0.220\pm0.012$ & $0.230\pm0.008$ & $0.415\pm0.006$ & $0.45\pm0.03$ & $0.28\pm0.04$ & $0.13\pm0.01$   & $33/18$\\
200 GeV &          & $p$      & $0.300\pm0.006$ & $0.310\pm0.008$ & $0.410\pm0.005$ & $0.51\pm0.04$ & $0.25\pm0.03$ & $0.053\pm0.003$ & $6/21$ \\
        &          & $\pi^-$  & $0.180\pm0.007$ & $0.205\pm0.008$ & $0.434\pm0.005$ & $0.75\pm0.02$ & $0.20\pm0.02$ & $0.99\pm0.10$   & $3/21$\\
        &          & $K^-$    & $0.184\pm0.009$ & $0.211\pm0.010$ & $0.409\pm0.005$ & $0.50\pm0.02$ & $0.28\pm0.03$ & $0.13\pm0.01$   & $1/18$\\
        &          &$\bar p$  & $0.281\pm0.006$ & $0.298\pm0.008$ & $0.368\pm0.005$ & $0.64\pm0.04$ & $0.11\pm0.01$ & $0.050\pm0.003$ & $16/21$\\
\cline{2-10}
        & 60--88\% & $\pi^+$  & $0.150\pm0.007$ & $0.313\pm0.005$ & $-$             & $0.84\pm0.02$ & $-$           & $0.21\pm0.02$   & $23/21$\\
        &          & $K^+$    & $0.150\pm0.008$ & $0.290\pm0.008$ & $0.384\pm0.003$ & $0.75\pm0.01$ & $0.10\pm0.01$ & $0.036\pm0.002$ & $13/18$\\
        &          & $p$      & $0.170\pm0.008$ & $0.260\pm0.006$ & $0.405\pm0.005$ & $0.30\pm0.02$ & $0.50\pm0.02$ & $0.015\pm0.001$ & $11/21$\\
        &          & $\pi^-$  & $0.155\pm0.007$ & $0.313\pm0.003$ & $-$             & $0.84\pm0.02$ & $-$           & $0.21\pm0.02$   & $17/21$\\
        &          & $K^-$    & $0.123\pm0.007$ & $0.205\pm0.005$ & $0.428\pm0.006$ & $0.57\pm0.02$ & $0.27\pm0.02$ & $0.035\pm0.002$ & $2/18$\\
        &          & $\bar p$ & $0.188\pm0.008$ & $0.293\pm0.012$ & $0.295\pm0.004$ & $0.40\pm0.03$ & $0.25\pm0.02$ & $0.011\pm0.001$ & $15/21$\\
\hline
Fig. 3   & 0--5\%   & $\pi^++\pi^-$ & $0.132\pm0.004$ & $0.284\pm0.004$ & $0.320\pm0.005$ & $0.62\pm0.03$ & $0.20\pm0.01$ & $240.00\pm14.00$& $5/38$\\
Pb-Pb    &          & $K^++K^-$     & $0.220\pm0.009$ & $0.321\pm0.006$ & $-$             & $0.28\pm0.06$ & $-$           & $36.00\pm1.50$  & $1/33$\\
2.76 TeV &          & $p+\bar p$    & $0.431\pm0.007$ & $0.454\pm0.014$ & $-$             & $0.85\pm0.15$ & $-$           & $9.70\pm0.30$   & $8/34$\\
\cline{2-10}
         & 70--80\% & $\pi^++\pi^-$ & $0.127\pm0.007$ & $0.136\pm0.005$ & $0.324\pm0.004$ & $0.38\pm0.02$ & $0.40\pm0.03$ & $5.60\pm0.27$   & $5/38$\\
         &          & $K^++K^-$     & $0.130\pm0.010$ & $0.267\pm0.009$ & $0.298\pm0.004$ & $0.17\pm0.04$ & $0.15\pm0.05$ & $0.82\pm0.04$   & $3/33$\\
         &          & $p+\bar p$    & $0.309\pm0.012$ & $0.365\pm0.013$ & $0.380\pm0.004$ & $0.10\pm0.04$ & $0.20\pm0.05$ & $0.30\pm0.02$   & $8/34$\\
\hline
Fig. 4   & 0--5\%   & $\pi^++\pi^-$ & $0.110\pm0.007$ & $0.351\pm0.012$ & $0.392\pm0.004$ & $0.73\pm0.01$ & $0.25\pm0.06$ & $6.37\pm0.37$   & $20/38$\\
$p$-Pb   &          & $K^++K^-$     & $0.162\pm0.012$ & $0.390\pm0.004$ & $-$             & $0.49\pm0.03$ & $-$           & $0.95\pm0.04$   & $3/28$\\
5.02 TeV &          & $p+\bar p$    & $0.325\pm0.016$ & $0.361\pm0.018$ & $0.470\pm0.007$ & $0.32\pm0.02$ & $0.42\pm0.08$ & $0.36\pm0.01$   & $10/36$\\
\cline{2-10}
         & 60--80\% & $\pi^++\pi^-$ & $0.120\pm0.008$ & $0.133\pm0.009$ & $0.342\pm0.006$ & $0.60\pm0.02$ & $0.22\pm0.02$ & $1.65\pm0.11$   & $86/38$\\
         &          & $K^++K^-$     & $0.158\pm0.009$ & $0.264\pm0.012$ & $0.341\pm0.004$ & $0.30\pm0.03$ & $0.40\pm0.04$ & $0.18\pm0.01$   & $28/28$\\
         &          & $p+\bar p$    & $0.218\pm0.004$ & $0.422\pm0.005$ & $0.475\pm0.006$ & $0.60\pm0.02$ & $0.22\pm0.02$ & $0.19\pm0.01$   & $23/36$\\
\hline\hline
\end{tabular}%
\end{center}}
\end{table*}

\begin{figure*}[htbp]
\begin{center}
\includegraphics[width=13.0cm]{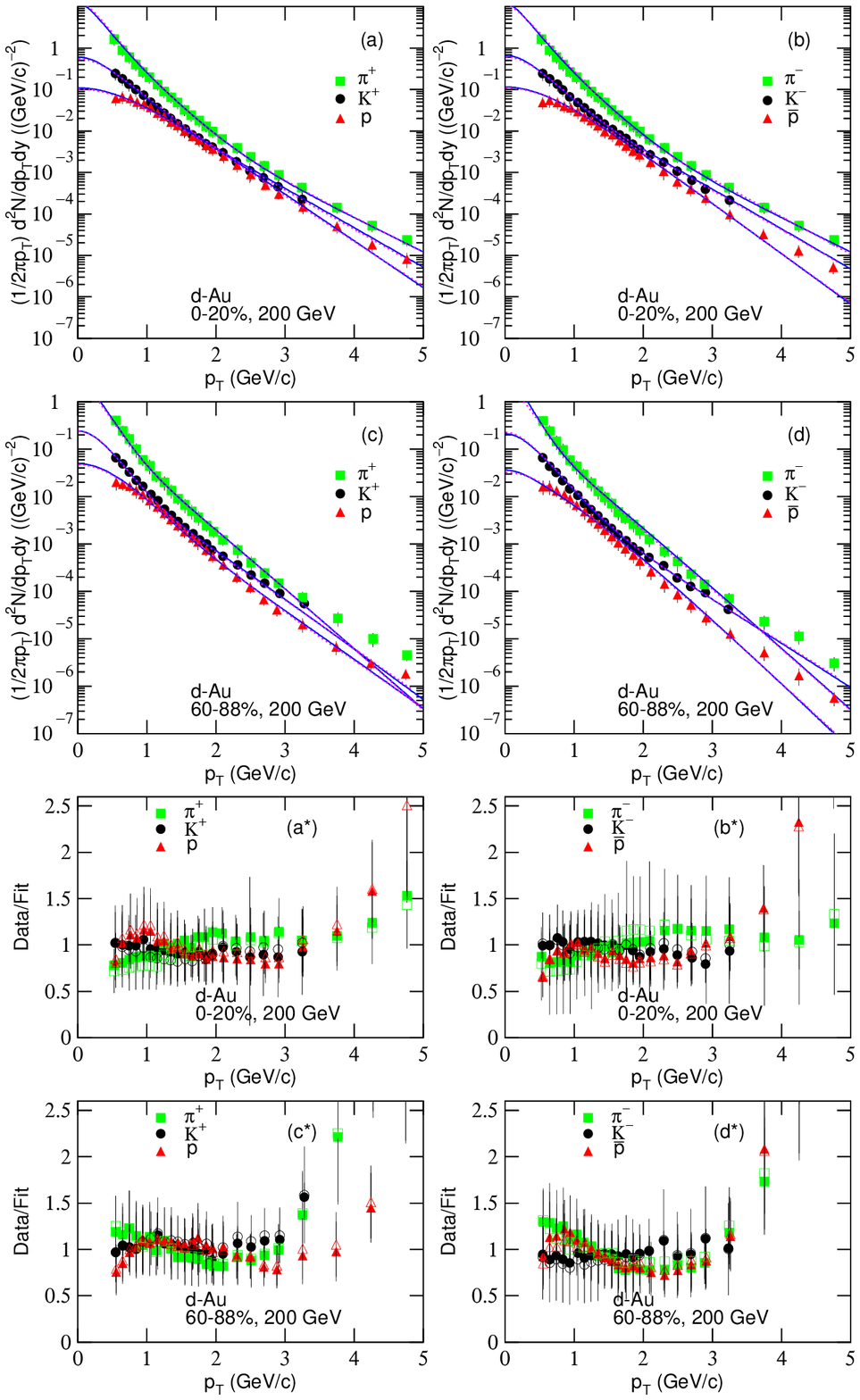}
\end{center}
{\small Fig. 2. Same as Fig. 1, but showing the spectra in (a)-(b)
0--20\% and (c)-(d) 60--88\% $d$-Au collisions at
$\sqrt{s_{NN}}=200$ GeV. The symbols represent the experimental
data measured by the PHENIX Collaboration in
$|\eta|<0.35$~\cite{12}.}
\end{figure*}

\begin{figure*}[!htb]
\begin{center}
\includegraphics[width=13.0cm]{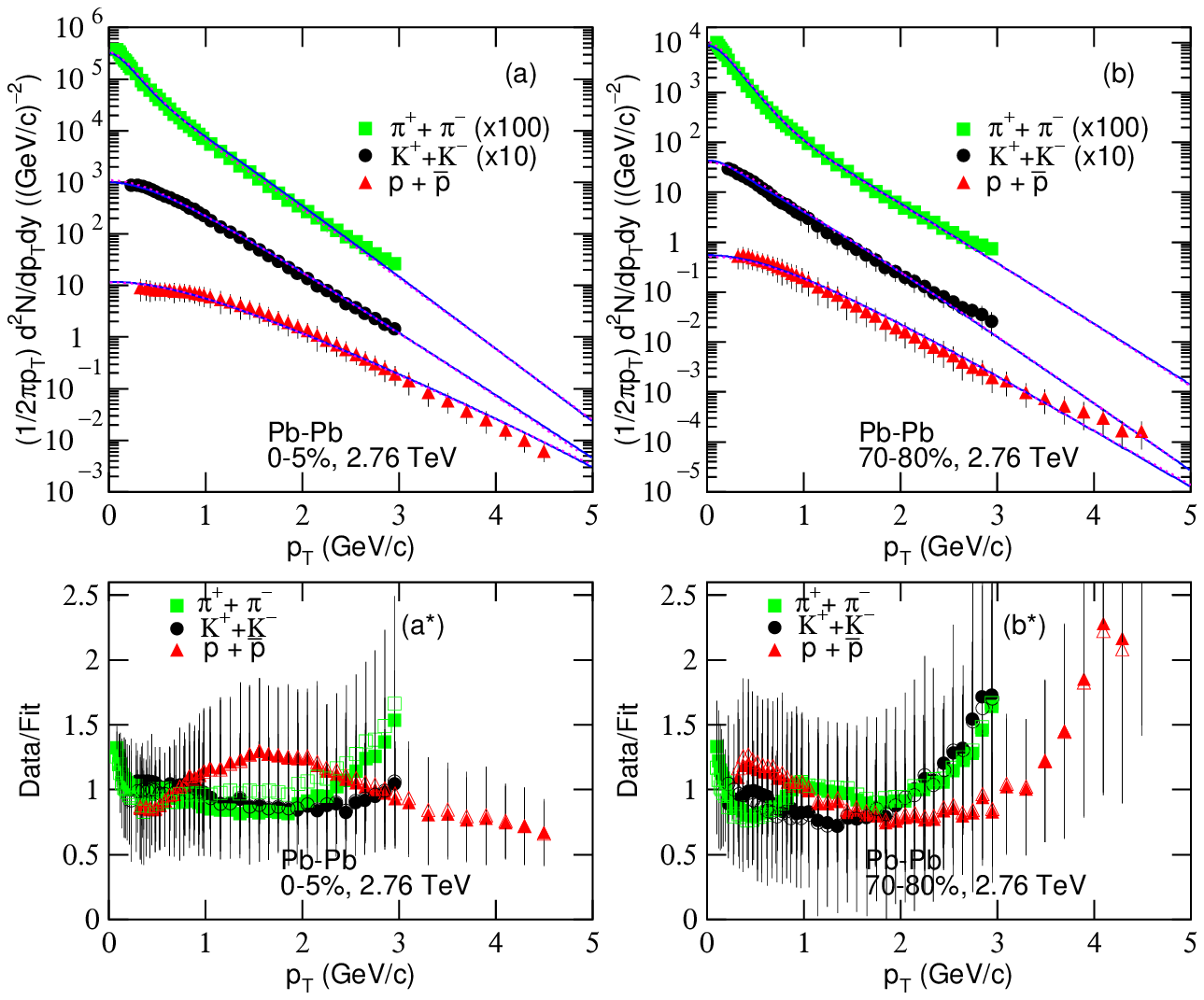}
\end{center}
{\small Fig. 3. Transverse momentum spectra of (a)-(b)
$\pi^++\pi^-$, $K^++K^-$, and $p+\bar p$ produced in (a) 0--5\%
and (b) 70--80\% Pb-Pb collisions at $\sqrt{s_{NN}}=2.76$ GeV,
where the spectra for different particles are multiplied by
different amounts shown in the panels for the clarity. The symbols
represent the experimental data measured by the ALICE
Collaboration in $|y|<0.5$~\cite{13,14}. The solid and dotted
curves are our results fitted by Eqs. (4) and (8) respectively.
The ratios of data/fit corresponding to panels (a) and (b) are
presented by panels (a$^*$) and (b$^*$) respectively, where the
closed and open symbols are the results due to Eqs. (4) and (8)
respectively.}
\end{figure*}

\begin{figure*}[!htb]
\begin{center}
\includegraphics[width=13.0cm]{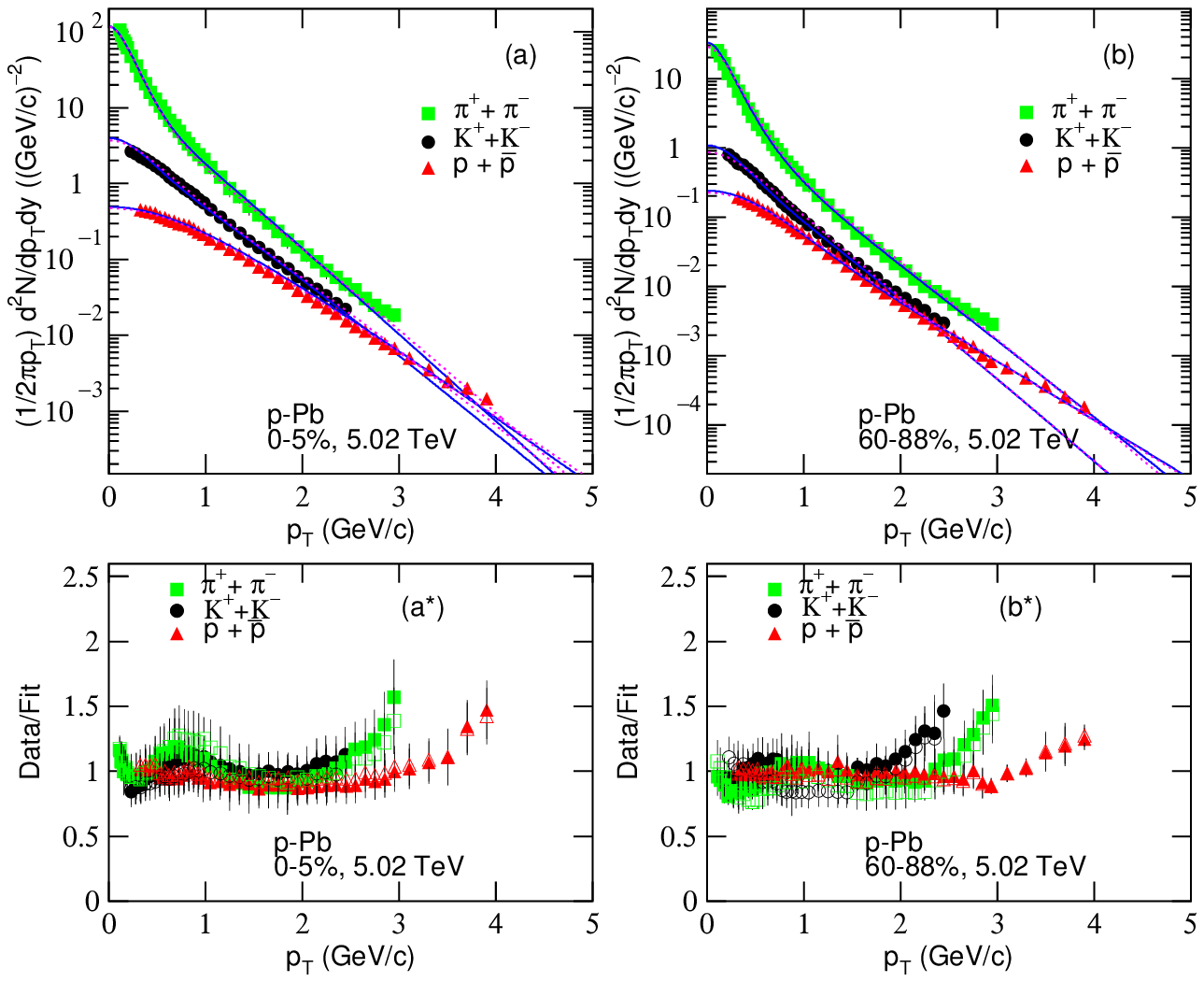}
\end{center}
{\small Fig. 4. Same as Fig. 3, but showing the spectra in (a)
0--5\% and (b) 60--88\% $p$-Pb collisions at $\sqrt{s_{NN}}=5.02$
GeV. The symbols represent the experimental data measured by the
ALICE Collaboration in $0<y<0.5$~\cite{15}.}
\end{figure*}

The transverse momentum spectra of (a)-(b) $\pi^++\pi^-$,
$K^++K^-$, and $p+\bar p$ produced in (a) 0--5\% and (b) 70--80\%
Pb-Pb collisions at $\sqrt{s_{NN}}=2.76$ TeV are displayed in Fig.
3. For the clarity, the spectra for $\pi^++\pi^-$ and $K^++K^-$
are multiplied by 100 and 10 respectively, which are shown in the
panels. The squares, circles, and triangles represent respectively
the experimental data of $\pi^++\pi^-$, $K^++K^-$, and $p+\bar p$
measured by the ALICE Collaboration in the rapidity range
$|y|<0.5$~\cite{13,14}. The solid and dotted curves are our
results fitted by the first component in Eq. (11) through Eqs. (4)
and (8) respectively. The related parameters are listed in Tables
1 and 2 with $\chi^2$/dof. The ratios of data/fit corresponding to
panels (a) and (b) are presented by panels (a$^*$) and (b$^*$)
respectively, where the closed and open symbols are the results
due to Eqs. (4) and (8) respectively. One can see that the two
models fit well the trends of the experimental data in low-$p_T$
region in Pb-Pb collisions at the LHC energy.

Figure 4 is the same as Fig. 3, but it shows the spectra in
(a)-(b) 0--5\% and (c)-(d) 60--88\% $p$-Pb collisions at
$\sqrt{s_{NN}}=5.02$ TeV. The symbols represent the experimental
data measured by the ALICE Collaboration in $0<y<0.5$~\cite{15}.
The curves are our fitted results and the related parameters are
listed in Tables 1 and 2 with $\chi^2$/dof. One can see that the
two models fit well the trends of the experimental data in
low-$p_T$ region in $d$-Au collisions at the top RHIC energy.

As effective temperatures, $T_1$ and $T_2$ depend on $m_0$, which
are shown in Figs. 5(a) and 5(b) respectively. Different symbols
shown in the panel represent the results from positive or negative
particles produced in central or peripheral Au-Au or $d$-Au
collisions, or from positive plus negative particles produced in
central or peripheral Pb-Pb or $p$-Pb collisions. These symbols
represent the results weighted different contribution fractions in
two or three-component listed in Tables 1 and 2. The dashed lines
are the fitting results by linear functions for negative particles
in Au-Au or $d$-Au collisions, and the solid lines are for other
cases. The intercepts and slopes of these linear functions are
listed in Table 3 with $\chi^2$, where ${\rm dof}=1$ is neglected.
Similar to Figs. 5(a) and 5(b), the dependences of $T_1$ and $T_2$
on centrality $C$ are displayed in Figs. 5(c) and 5(d)
respectively. The different symbols represent different $T_1$
($T_2$) for different cases shown in the panels, and $T_1$ ($T_2$)
obtained from the spectra of positive and negative particles are
not distinguished to avoid trivialness. One can see that the
effective temperature obtained from the spectrum of particles with
large mass is obviously larger than that with small mass. The
effective temperature in central collisions is slightly larger
than or equal to that in peripheral collisions. The effective
temperature at LHC energy is comparable with that at RHIC energy.
Meanwhile, the effective temperature in Au-Au (Pb-Pb) collisions
is comparable with that in $d$-Au ($p$-Pb) collisions.

We can regarded the intercepts in the linear relation between the
effective temperature and particle mass in Fig. 5(a) and 5(b) as
the kinetic freeze-out temperatures $T_{01}$ and
$T_{02}$~\cite{10,11} respectively. Figures 6(a) and 6(b) show the
dependences of $T_{01}$ and $T_{02}$ on $C$ respectively, where
the symbols represent the intercepts obtained from Figs. 5(a) and
5(b) and listed in Table 3. One can see that the kinetic
freeze-out temperature in central collisions is slightly larger
than or equal to that in peripheral collisions. The kinetic
freeze-out temperatures in collisions at LHC and RHIC energies are
comparable with each other. Meanwhile, the kinetic freeze-out
temperatures in Au-Au (Pb-Pb) and $d$-Au ($p$-Pb) collisions are
comparable with each other. These results confirm our previous
works~\cite{29,30}, though the absolute values are different due
to different methods. These results are also in agreement with the
effective temperature in general trends in different centralities,
at different energies, and in different collision systems.

\begin{table*}[!htb]
{Table 3. Values of intercepts, slopes, and $\chi^2$ in the linear
fittings in Fig. 5, where ${\rm dof}=1$ is not shown in the table
to avoid trivialness. \vspace{-.3cm}
\begin{center}
\begin{tabular}{ccccccccccc}\\ \hline\hline
Figure & Collisions & Centrality & Particle &
Intercept (GeV) & Slope ($c^2$) & $\chi^2$\\
\hline
Fig. 5(a) & Au-Au  & Central    & positive          & $0.294\pm0.004$ & $0.425\pm0.005$ & $13$\\
          &        &            & negative          & $0.385\pm0.003$ & $0.294\pm0.004$ & $8$ \\
          &        & Peripheral & positive          & $0.230\pm0.005$ & $0.331\pm0.005$ & $47$\\
          &        &            & negative          & $0.232\pm0.004$ & $0.394\pm0.006$ & $14$\\
          & $d$-Au & Central    & positive          & $0.371\pm0.004$ & $0.335\pm0.003$ & $1$ \\
          &        &            & negative          & $0.386\pm0.004$ & $0.243\pm0.005$ & $2$ \\
          &        & Peripheral & positive          & $0.370\pm0.004$ & $0.144\pm0.003$ & $1$ \\
          &        &            & negative          & $0.345\pm0.003$ & $0.221\pm0.003$ & $1$ \\
          & Pb-Pb  & Central    & positive+negative & $0.361\pm0.008$ & $0.631\pm0.005$ & $23$\\
          &        & Peripheral & positive+negative & $0.310\pm0.005$ & $0.449\pm0.005$ & $2$ \\
          & $p$-Pb & Central    & positive+negative & $0.363\pm0.008$ & $0.585\pm0.003$ & $1$ \\
          &        & Peripheral & positive+negative & $0.317\pm0.005$ & $0.374\pm0.005$ & $11$\\
\hline
Fig. 5(b) & Au-Au  & Central    & positive          & $0.235\pm0.002$ & $0.110\pm0.005$ & $1$ \\
          &        &            & negative          & $0.182\pm0.002$ & $0.136\pm0.003$ & $1$\\
          &        & Peripheral & positive          & $0.168\pm0.002$ & $0.113\pm0.003$ & $1$ \\
          &        &            & negative          & $0.167\pm0.003$ & $0.092\pm0.004$ & $3$ \\
          & $d$-Au & Central    & positive          & $0.215\pm0.003$ & $0.124\pm0.003$ & $1$ \\
          &        &            & negative          & $0.167\pm0.003$ & $0.144\pm0.004$ & $1$ \\
          &        & Peripheral & positive          & $0.161\pm0.003$ & $0.106\pm0.002$ & $2$ \\
          &        &            & negative          & $0.166\pm0.003$ & $0.092\pm0.003$ & $3$ \\
          & Pb-Pb  & Central    & positive+negative & $0.158\pm0.003$ & $0.291\pm0.004$ & $1$ \\
          &        & Peripheral & positive+negative & $0.136\pm0.002$ & $0.252\pm0.005$ & $1$ \\
          & $p$-Pb & Central    & positive+negative & $0.140\pm0.002$ & $0.261\pm0.004$ & $1$ \\
          &        & Peripheral & positive+negative & $0.135\pm0.004$ & $0.185\pm0.004$ & $3$ \\
\hline\hline
\end{tabular}%
\end{center}}
\end{table*}

\begin{figure*}[!htb]
\begin{center}
\vskip0.5cm
\includegraphics[width=16.0cm]{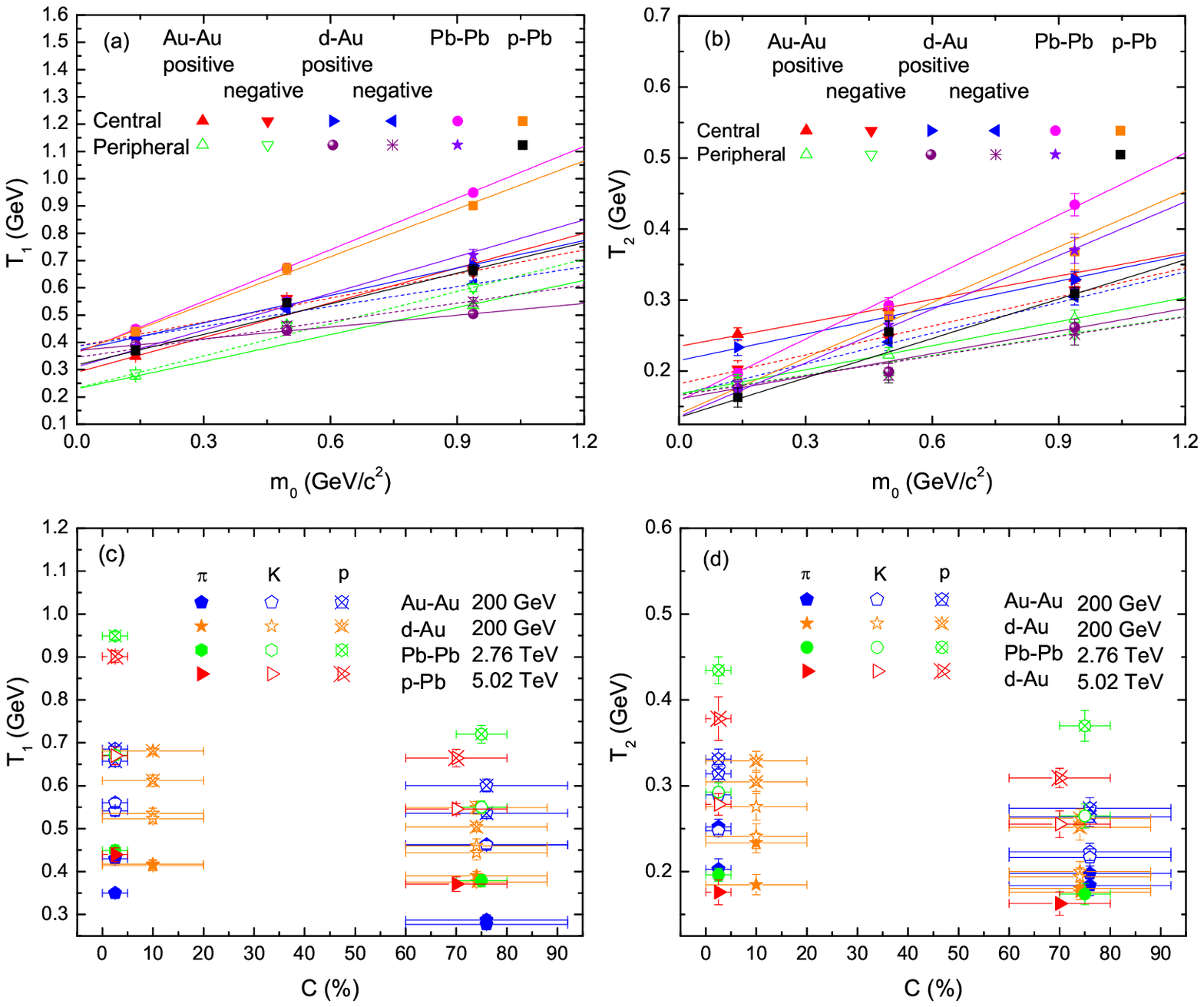}
\end{center}
{\small Fig. 5. Upper panel: Dependences of (a) $T_1$ and (b)
$T_2$ on $m_0$. Different symbols represent the results from
positive or negative particles produced in central or peripheral
Au-Au or $d$-Au collisions, or from positive plus negative
particles produced in central or peripheral Pb-Pb or $p$-Pb
collisions. The dashed lines are the fitting results by linear
functions for negative particles in Au-Au or $d$-Au collisions,
and the solid lines are for other cases. Lower panel: Dependences
of (c) $T_1$ and (d) $T_2$ on $C$. The symbols are not
distinguished for positive and negative particles in Au-Au and
$d$-Au collisions, and the positive and negative particles are
considered together in Pb-Pb and $p$-Pb collisions.}
\end{figure*}

\begin{figure*}[!htb]
\begin{center}
\vskip0.5cm
\includegraphics[width=16.0cm]{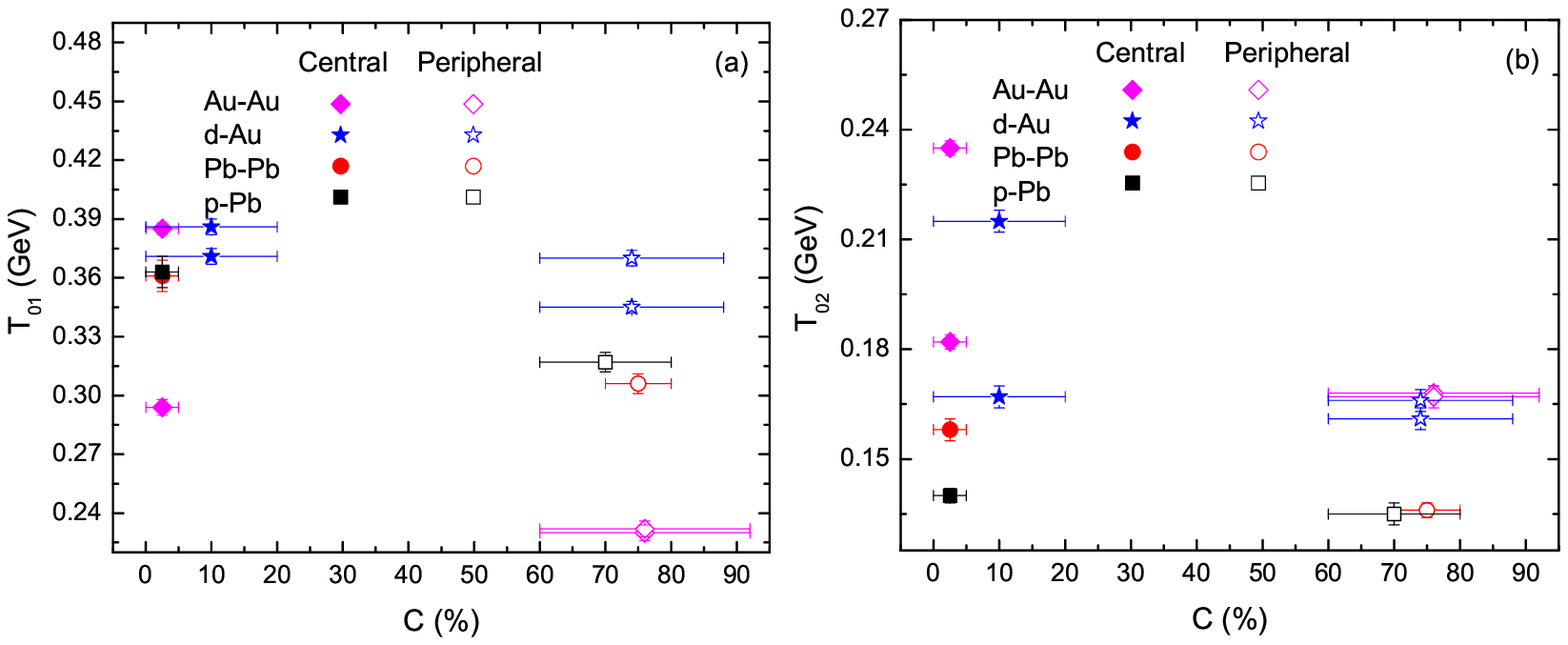}
\end{center}
{\small Fig. 6. Dependences of (a) $T_{01}$ and (b) $T_{02}$ on
$C$. Different symbols represent the results from positive or
negative particles produced in central or peripheral Au-Au or
$d$-Au collisions, or from positive plus negative particles
produced in central or peripheral Pb-Pb or $p$-Pb collisions. The
symbols are not distinguished for positive and negative particles
in Au-Au and $d$-Au collisions, and the positive and negative
particles are considered together in Pb-Pb and $p$-Pb collisions.}
\end{figure*}

\begin{figure*}[!htb]
\begin{center}
\vskip0.5cm
\includegraphics[width=16.0cm]{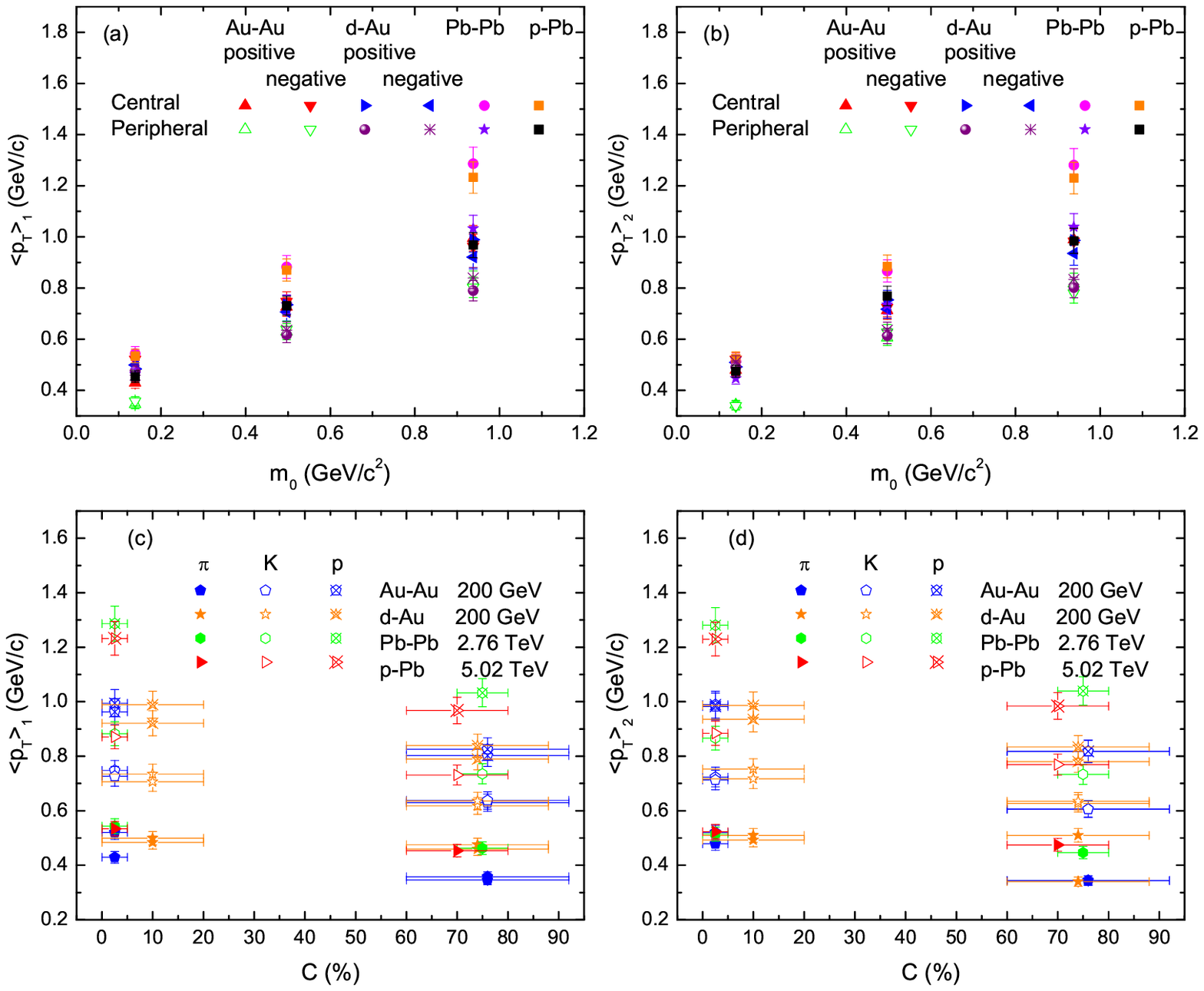}
\end{center}
{\small Fig. 7. Upper panel: Dependences of (a) $\langle
p_T\rangle_1$ and (b) $\langle p_T\rangle_2$ on $m_0$. Different
symbols represent the results from positive or negative particles
produced in central or peripheral Au-Au or $d$-Au collisions, or
from positive plus negative particles produced in central or
peripheral Pb-Pb or $p$-Pb collisions. Lower panel: Dependences of
(c) $\langle p_T\rangle_1$ and (d) $\langle p_T\rangle_2$ on $C$.
The symbols are not distinguished for positive and negative
particles in Au-Au and $d$-Au collisions, and the positive and
negative particles are considered together in Pb-Pb and $p$-Pb
collisions.}
\end{figure*}

\begin{figure*}[!htb]
\begin{center}
\vskip0.5cm
\includegraphics[width=16.0cm]{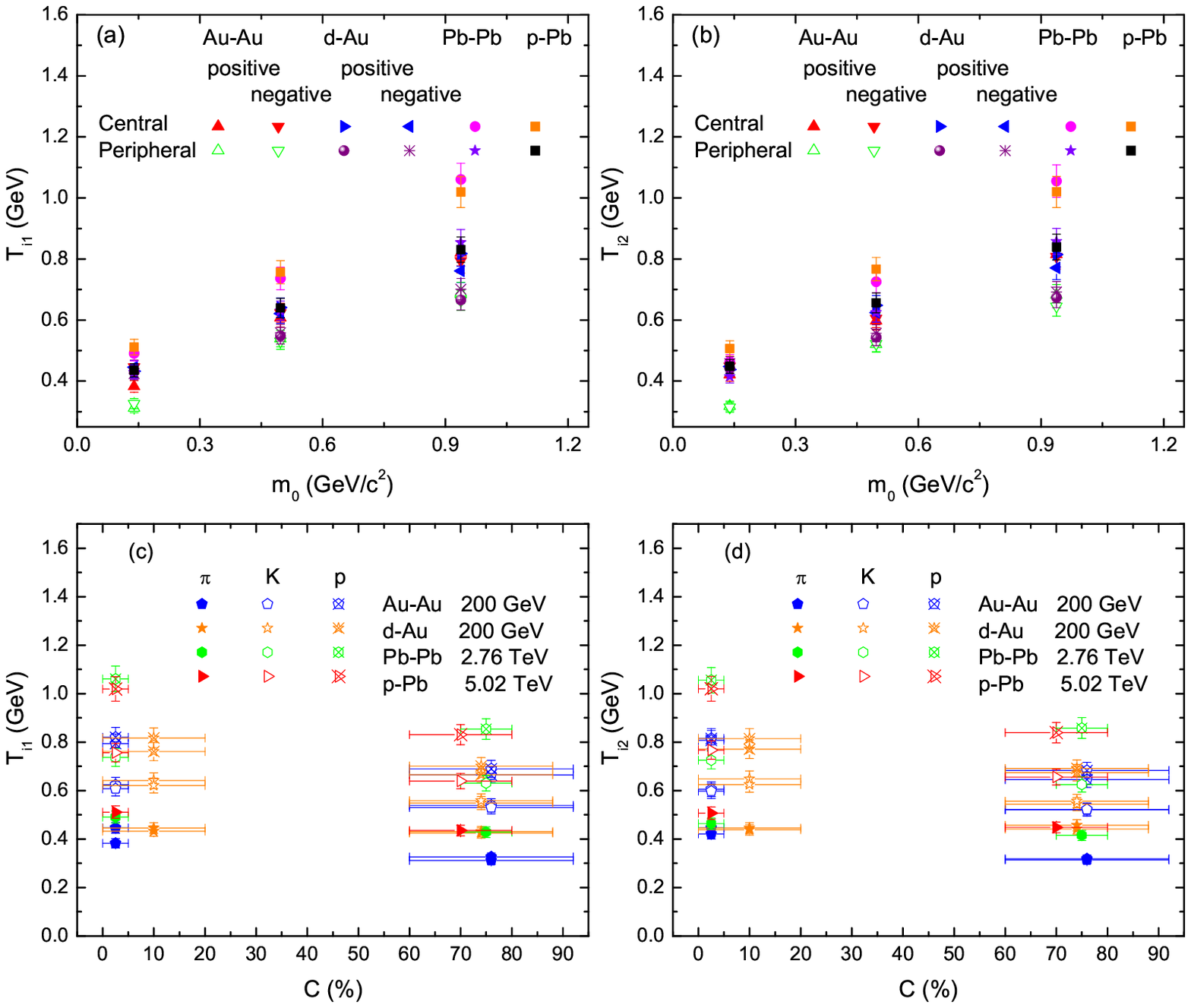}
\end{center}
{\small Fig. 8. The same as Fig. 7, but showing the dependences of
(a) $T_{i1}$ and (b) $T_{i2}$ on $m_0$, as well as the dependences
of (c) $T_{i1}$ and (d) $T_{i2}$ on $C$.}
\end{figure*}

Figures 7(a) and 7 (b) show the dependences of mean $p_T$
($\langle p_T \rangle_1$ and $\langle p_T \rangle_2$) on $m_0$
obtained from the fits of multi-component Hagedorn thermal model
and standard distribution respectively. Figures 7(c) and 7 (d)
show the dependences of $\langle p_T \rangle_1$ and $\langle p_T
\rangle_2$ on $C$ respectively. The symbols represent $\langle p_T
\rangle_1$ and $\langle p_T \rangle_2$ obtained from the fitting
functions (Eqs. (4) and (8)) and their parameter values (Tables 1
and 2) over a $p_T$ range from 0 to 5 GeV/$c$. The results
obtained from the spectra of positive and negative particles are
not distinguished to avoid trivialness. One can see that the mean
$p_T$ for particles with large mass, in central collisions, at LHC
energy, and in Au-Au (Pb-Pb) collisions are larger than or equal
to those for particles with small mass, in peripheral collisions,
at RHIC energy, and in $d$-Au ($p$-Pb) collisions, respectively.
These results are in agreement with the trends of effective and
kinetic freeze-out temperatures in different centralities, at
different energies, and for different system sizes. These results
are also in agreement with the trend of effective temperature for
emissions of particles with different masses.

Figure 8 is the same as Fig. 7, but it shows the dependences of
initial temperature (a)(c) $T_{i1}$ and (b)(d) $T_{i2}$ on (a)(b)
$m_0$ and (c)(d) $C$, where $T_{i1}$ ($T_{i2}$) is obtained by the
root-mean-square $p_T$ divided by $\sqrt{2}$, i.e. $\sqrt{\langle
p^2_T \rangle_1/2}$ ($\sqrt{\langle p^2_T \rangle_2/2}$) according
to~\cite{31,32,33}. The symbols represent the results obtained
from the fitting functions (Eqs. (4) and (8)) and their parameter
values (Tables 1 and 2) over a $p_T$ range from 0 to 5 GeV/$c$.
One can see that the trends of $T_{i1}$ ($T_{i2}$) on particle
mass, event centrality, collision energy, and system size are
similar to those of $\langle p_T \rangle_1$ ($\langle p_T
\rangle_2$), though $T_{i1}$ ($T_{i2}$) is smaller than $\langle
p_T \rangle_1$ ($\langle p_T \rangle_2$).

In the rectangular coordinate system which regards the collision
point as the original $O$, one of the beam directions as the $Oz$
axis, and the reaction plane as the $xOz$ plane, the
root-mean-square $p_T$ divided by $\sqrt{2}$, i.e. $\sqrt{\langle
p^2_T \rangle/2}$ also equals to the root-mean-square momentum
component ($\sqrt{\langle p^2_x \rangle}$ or $\sqrt{\langle p^2_y
\rangle}$) in the transverse plane $xOy$. In the rest frame of an
isotropic emission source, $\sqrt{\langle p^2_T \rangle/2}$ also
equals to $\sqrt{\langle p^2_z \rangle}$. Thus, Fig. 8 also
reflects the trends of $\sqrt{\langle p^2_x \rangle}$ and
$\sqrt{\langle p^2_y \rangle}$ in different conditions. In the
rest frame of an isotropic emission source, Fig. 8 also reflects
the trends of $\sqrt{\langle p^2_z \rangle}$ in different
conditions.

From Figs. 5--8, one can see that the initial temperature is
larger than the effective temperature, and the latter is larger
than the kinetic freeze-out temperature. Generally, the chemical
freeze-out temperature is between the initial and kinetic
freeze-out temperatures, and is approximately equal to the
effective temperature. This order is in agreement with the order
of time evolution of the interacting system, though both the
effective and kinetic freeze-out temperatures are extracted at the
kinetic freeze-out. However, we cannot compare directly the four
temperatures in most case due to different thermometric scales
being used. Like the thermometric scales used in thermal and
statistical physics, we need a method to unify different
thermometric scales in subatomic physics. To structure the method
is beyond the focus of the present work. We shall not discuss this
issue in the present work. Or, we may use the quantities which are
model independent to describe the temperatures.

The results presented in Figs. 5--8 also reflect that the violent
degree of impact and squeeze in central collisions is comparable
with that in peripheral collisions. Meanwhile, the violent degree
of impact and squeeze in collisions at LHC energy is comparable
with that at RHIC energy. These results are natural due to the
fact that the amount of energy deposited in central collisions is
comparable with that in peripheral collisions. Meanwhile, the
amount of energy deposited in collisions at LHC energy is
comparable with that at RHIC energy. The similar results in Au-Au
(Pb-Pb) and $d$-Au ($p$-Pb) collisions confirms the effect of the
heaviest nucleus~\cite{30} which states that the excitation degree
depends on the heaviest nucleus, but not the lightest nucleus and
total number of nucleons, in proton(deuteron)-nucleus and
nucleus-nucleus collisions at a given $\sqrt{s_{NN}}$.

In the above discussions on Figs. 1--8, we have not mentioned the
(pseudo)critical temperature at equilibrium, though it is an
important baseline in this context. In fact, to obtain
experimentally the critical temperature from the fits to data, we
have to study the energy dependent parameters, though the critical
temperature can be calculated from first principles such as
lattice QCD~\cite{33a}. Based on an experimental point of view, it
is known that there is few related studies on the initial and
effective temperatures in literature. In principle, the critical
temperature at chemical (kinetic) freeze-out can be obtained from
experimental excitation function of chemical (kinetic) freeze-out
temperature. However, there is no critical energy being determined
in experiments at present. This renders that the critical
temperature at chemical (kinetic) freeze-out is not determined in
experiments.

Theoretically, the critical temperature at chemical freeze-out is
model dependent in some cases, though it is usually less than 170
MeV~\cite{34}. For examples, the thermal and statistical model
proposes the limiting or approximate critical temperature being
about 160 MeV~\cite{2,3,4,5,34}. The calculation based on
hydrodynamic evolution proposes the critical temperature being 155
MeV~\cite{35}. The theory based on lattice QCD shows the critical
temperature being around 160 MeV~\cite{36,37,38,38a,38b,38c}, or
as low as 140 MeV~\cite{38d,38e}, which comes from first
principles and should be model independent. The analysis based on
finite size scaling and hydrodynamics shows the critical
temperature is about 165 MeV~\cite{38f,38g}. The result based on a
realistic Polyakov--Nambu--Jona-Lasinio (rPNJL) model gives the
critical temperature is 93 MeV~\cite{39}. Comparing with the
indeterminacy of critical temperature at chemical freeze-out, the
study of critical temperature at kinetic freeze-out is lacking in
literature.

In consideration of lattice QCD deriving the critical temperature
from first principles~\cite{33a,36,37,38,38a,38b,38c,38d,38e}
which is model independent, other models should regard lattice QCD
as a standard to revise their calculations to coordinate with the
standard. In fact, some models such as the thermal and statistical
model~\cite{2,3,4,5,34}, the calculation based on hydrodynamic
evolution~\cite{35}, and the analysis based on finite size scaling
and hydrodynamics~\cite{38f,38g} are approximately harmonious with
lattice QCD with small differences. The rPNJL model~\cite{39}
studies the critical endpoint which is a different observable
which appears at finite baryon density, this is why the critical
temperature (93 MeV) there is so low.

From the point of view in physics meaning, the temperatures
discussed in Lattice QCD and other related models are transition
temperature, which are different from initial, effective, and
kinetic freeze-out temperatures discussed in the present work. The
three temperatures extracted from the multi-component Hagedorn
thermal model (standard distribution) are harmonious in relative
size. The multi-component Hagedorn thermal model results in larger
temperatures than the multi-component standard distribution. The
kinetic freeze-out temperature extracted from the multi-component
standard distribution is approximately harmonious with the
transition temperature discussed in Lattice QCD in order of size.

Based on the thermal and statistical model, the excitation
function of chemical freeze-out temperature obtained from particle
ratios at mid-rapidity in central nucleus-nucleus collisions show
a quickly increase at a few GeV, a slowly increase at about 10
GeV, and then a saturation at RHIC and LHC
energies~\cite{2,3,4,5,34}. The limiting temperature at saturation
is $\sim$164 MeV~\cite{2,3,4,5} or a little less ($\sim$158
MeV~\cite{34}). Other models~\cite{35,36,37,38,39} are expected to
show similar excitation functions of chemical freeze-out
temperature with different limiting values, though most excitation
functions are lacking in literature. With the limiting or
approximate critical temperatures, the limiting baryon chemical
potentials are 1 MeV or 0, and the critical baryon chemical
potentials obtained from different
models~\cite{2,3,4,5,34,35,36,37,38,39} are in a wide range from
95 to 720 MeV which are also model dependent in some cases.

In the framework of the blast-wave model or similar model,
considering different velocities of transverse flow, the
excitation function of kinetic freeze-out temperature obtained
from transverse momentum spectra at mid-rapidity in central
nucleus-nucleus collisions also show a quickly increase at a few
GeV, a slowly change at about 10 GeV, and then a
saturation~\cite{13,40} or a slight increase~\cite{29,30} or a
continuous decrease~\cite{41} or a decrease till the top RHIC
energy and then invariability till the LHC
energies~\cite{42,43,44,45,46}. The alternative method show a
slight increase from the top RHIC energy to the LHC
energies~\cite{29,30}. In deeded, the trend of kinetic freeze-out
temperature at high energy is model dependent. As a parameter
entangled to kinetic freeze-out temperature, the transverse flow
velocity also show a model dependent trend with collision energy.

Before summary and conclusions, we would like to emphasize that
the significance of the present work is not solely to describe the
$p_T$ spectra themselves, but mainly to extract various
temperatures via the description of the $p_T$ spectra. Although
the hydrodynamic models are excellent, the thermal and statistical
models are also useful. In particular, the statistical method is
more closer to experiments themselves. We fit the $p_T$ spectra by
taking a sum of two or three statistical distributions after
integrating over the necessary rapidity window, in which the
contribution of resonance decays at low $p_T$ is naturally
described by the first distribution. Once in a while, an inverse
power-law is also added to account for high $p_T$ part of the
data.

The inverse power-law, i.e. the second item in Eqs. (10) and (11),
has no contribution to temperatures. To show clearly the trend of
the first item in Eqs. (10) and (11), the contribution of the
second item therein is not included, which results in the same
result from Eqs. (10) and (11) and the fits to be poor at large
$p_T$ in some cases in Figs. 1--4. Although the curvatures in the
data are a result of a complex and rich interplay of temperature,
flow, and hard scattering process of hadronization as well as
decay contributions and even viscosity, the present work discusses
an alternative method to extract various temperatures from the
contaminative data as accurately as possible.

The values of effective and kinetic freeze-out temperatures
obtained in the present work have only the relative significance
due to the fact that the measured functions which can be regarded
as ``thermometers" or ``thermometric scales" used in the present
work are different from others such as that of chemical freeze-out
temperature which is based on the ratios of different types of
particles in thermal and statistical model~\cite{2,3,4,5,34}. In
fact, ``thermometers" or ``thermometric scales" used in subatomic
physics should be unified in the framework of standard
distribution which is equal to or closest to the Boltzmann
distribution in thermal physics.

In particular, the initial temperature $T_i$ obtained in the
present work is model independent, though it is calculated from
the fitted curve. In fact, $T_i$ can be directly obtained from
data themselves if the data points having no large statistical
fluctuation. In addition, as a model independent quantity, the
half of mean transverse momentum $\langle p_T\rangle/2$ is
expected to represent the sum of contributions of thermal motion
and transverse flow, where $1/2$ is used due to both contributions
of projectile and target participants. Let $k$ denote the fraction
in $\langle p_T\rangle/2$ to contribute to $T_0$, we may define
$T_0\equiv k\langle p_T\rangle/2$ and $\beta_T\equiv (1-k)\langle
p_T\rangle/2\overline{m}$, where $\overline{m}$ denote the mean
energy (mean moving mass) of the considered particles in the
source rest frame in which particles are assumed to emit
isotropically~\cite{29,30}.

The new definition of $T_0$ ($\beta_T$) is model independent.
Contrastively, other definitions or extractions of $T_0$
($\beta_T$) are model dependent. In particular, in some cases, the
values of $T_0$ ($\beta_T$) obtained by other methods at high
energy are inconsistent due to different given
conditions~\cite{13,29,30,40,41,42,43,44,45,46}. To obtain
consistent results with the alternative method~\cite{29,30}, one
can revise the given conditions such as the profile function of
transverse flow in the blast-wave model~\cite{6,7,8,9}. Go a step
further, to avoid model dependence, we hope to study $T_0$ and
$\beta_T$ and their excitation functions from model independent
$\langle p_T\rangle$ in future.
\\

{\section{Summary and conclusions}}

We summarize here our main observations and conclusions.

(a) The transverse momentum spectra of $\pi^+$, $\pi^-$, $K^+$,
$K^-$, $p$, and $\bar p$ produced in central and peripheral Au-Au
and $d$-Au collisions at the top RHIC energy, as well as in
central and peripheral Pb-Pb and $p$-Pb collisions at LHC
energies, have been analyzed by the Hagedorn thermal model and the
standard distribution in terms of multi-component. The modelling
results are in agreement with the experimental data in low-$p_T$
region measured by the PHENIX Collaboration at the RHIC and by the
ALICE Collaboration at the LHC. The initial, effective, and
kinetic freeze-out temperatures are then extracted from the
fitting to the transverse momentum spectra.

(b) The initial temperature is larger than the effective
temperature, and the latter is larger than the kinetic freeze-out
temperature. The chemical freeze-out temperature is between the
initial and effective temperatures, and is approximately equal to
the effective temperature. This order is in agreement with the
order of time evolution of the interacting system, though both the
effective and kinetic freeze-out temperatures are extracted at the
kinetic freeze-out.

(c) The initial, effective, and kinetic freeze-out temperatures in
central collisions are respectively comparable with those in
peripheral collisions. The three types of temperatures at LHC
energy are respectively comparable with those at RHIC energy.
Although the three types of temperatures are different in values,
they show similar trends in different centralities and at
different energies.

(d) The violent degree of impact and squeeze in central collisions
is comparable with that in peripheral collisions, and the violent
degree of impact and squeeze in collisions at LHC energy is
comparable with that at RHIC energy. These results are caused due
to the amount of energy deposited in central collisions being
comparable with that in peripheral collisions, and the amount of
energy deposited in collisions at LHC energy being comparable with
that at RHIC energy.

(e) To use a model independent quantity to represent the kinetic
freeze-out temperature $T_0$ and the transverse flow velocity
$\beta_T$, we propose to define $T_0\equiv k\langle p_T\rangle$
and $\beta_T\equiv (1-k)\langle p_T\rangle/\overline{m}$, where
$k$ may be an energy dependent parameter which is needed to study
further. Considering $T_i\equiv \sqrt{\langle p_T^2\rangle/2}$
which is also model independent, one has three model independent
quantities, $T_i$, $T_0$, and $\beta_T$. As for the effective
temperature, we would like to give it up due to its model
dependence and ``non-real" one.
\\
\\
{\bf Data Availability}

The data used to support the findings of this study are included
within the article and are cited at relevant places within the
text as references.
\\
\\
{\bf Compliance with Ethical Standards}

The authors declare that they are in compliance with ethical
standards regarding the content of this paper.
\\
\\
{\bf Conflict of Interest}

The authors declare that there are no conflicts of interest
regarding the publication of this paper. The funders had no role
in the design of the study; in the collection, analysis, or
interpretation of the data; in the writing of the manuscript, or
in the decision to publish the results.
\\
\\
{\bf Acknowledgments}

Communications from Ashfaq Ahmad are highly acknowledged. This
work was supported by the National Natural Science Foundation of
China under Grant Nos. 11575103 and 11847311, the Chinese
Government Scholarship (China Scholarship Council), the Scientific
and Technological Innovation Programs of Higher Education
Institutions in Shanxi (STIP) under Grant No. 201802017, the
Shanxi Provincial Natural Science Foundation under Grant No.
201701D121005, and the Fund for Shanxi ``1331 Project" Key
Subjects Construction.
\\

\end{multicols}
\end{document}